\documentclass[aps,prl,twocolumn,superscriptaddress,showpacs]{revtex4}
\usepackage{graphicx}
\usepackage{amsmath}

\begin{document}

\newcommand{\IMS}{Institute for Microstructural Sciences, National
    Research Council of Canada, Ottawa, Ontario, Canada K1A 0R6}
\newcommand{\UofO}{Department of Physics, University of Ottawa, Ottawa,
    Ontario, Canada K1N 6N5}

\title{Voltage Induced Hidden Symmetry and Photon Entanglement Generation \\in a Single, Site-Selected Quantum Dot}


\author{M. E. Reimer}
    \affiliation{\IMS}
    \affiliation{\UofO}
\author{M. Korkusi\'{n}ski}
    \affiliation{\IMS}
\author{J. Lefebvre}
    \affiliation{\IMS}
\author{J. Lapointe}
   \affiliation{\IMS}
\author{P. J. Poole}
   \affiliation{\IMS}
\author{G. C. Aers}
   \affiliation{\IMS}
\author{D. Dalacu}
   \affiliation{\IMS}
\author{W. R. McKinnon}
   \affiliation{\IMS}
\author{S. Fr\'{e}d\'{e}rick}
    \affiliation{\IMS}
    \affiliation{\UofO}
\author{P. Hawrylak}
    \affiliation{\IMS}
    \affiliation{\UofO}
\author{R. L. Williams}
    \affiliation{\IMS}
    \affiliation{\UofO}

\email[]{Michael.Reimer@nrc.ca}


\date{\today}

\begin{abstract}
Present proposals for the realisation of entangled photon pair
sources using the radiative decay of the biexciton in semiconductor
quantum dots are limited by the need to enforce degeneracy of the
two intermediate, single exciton states. We show how this
requirement is lifted if the biexciton binding energy can be tuned
to zero and we demonstrate this unbinding of the biexciton in a
single, pre-positioned InAs quantum dot subject to a lateral
electric field. Full Configuration-Interaction calculations are
presented that reveal how the biexciton is unbound through
manipulation of the electron-hole Coulomb interaction and the
consequent introduction of Hidden Symmetry.
\end{abstract}


\maketitle

\newcommand{\bra}[1]{\langle #1|}
\newcommand{\ket}[1]{|#1\rangle}
\newcommand{\braket}[2]{\langle #1|#2\rangle}

Site-selected InAs/InP quantum dots emitting in the wavelength range
between 1300~nm and 1550~nm offer a fully scalable route to sources
of single photons and entangled photon pairs for fibre-based quantum
cryptography \cite{Shields07, Santori02, Gisin02, Ben00, Ekert91},
quantum teleportation \cite{Bennett93} and quantum repeater
technologies \cite{Gisin07}. In such strongly confined systems, the
Coulomb interactions between optically generated electrons and
holes, as well as the dot size and compositional structure, play a
major role in determining transition energies and oscillator
strengths. Presently, in order to produce polarisation entangled
photon pairs, the biexciton radiative decay cascade in single
semiconductor quantum dots (QDs) \cite{Ben00, Stevenson06,
Akopian06} is utilized. Entanglement in these schemes is destroyed
however if the degeneracy of the two possible spin configurations of
the intermediate neutral exciton ground state is lifted, as is
frequently the case in the presence of the electron-hole exchange
interaction in anisotropic quantum dots \cite{Bayer02}. As such,
there have been extensive efforts to remove the anisotropic exchange
splitting (AES) through the application of external magnetic
\cite{Stevenson06} and electric \cite{Kowalik05, Geradot07} fields
or through spectral filtering \cite{Akopian06} and quantum dot size
and composition engineering \cite{Gre06}.

In this Letter, we propose an alternative scheme for entangled
photon pair generation from quantum dots (QDs) that is robust to the
presence of anisotropic exchange splitting. By precisely balancing
the competing Coulomb attractions and repulsions for excitons
(\emph{X}) and biexcitons (\emph{XX}) within a single InAs/InP QD
using an applied lateral electric field, we enforce Hidden Symmetry
\cite{Haw03, Haw99, Bay00} within the dot. Under such circumstances,
the binding energy of the biexciton vanishes and it is possible to
identify two indistinguishable recombination pathways for the
biexciton, even in the presence of non-degenerate intermediate
exciton states. The observed behavior allows us to identify a
voltage controlled route to entangled photon pair generation even
from imperfect QDs, without resort to large external magnetic
fields, or materials engineering of individual dots.

Fig. 1 shows a schematic view of the sample construction used here
to apply a lateral electric field to single InAs/InP QDs. Dots are
nucleated at the apex of a "stripe geometry" InP nanotemplate and
subsequently capped with InP. Such templates, produced \emph{in
situ} during crystal growth, allow one to control the surface
migration of deposited InAs, so that the nucleation site of the dots
can be selected \emph{a} \emph{priori} \cite{Lefebvre02}. Pairs of
metallic, Schottky gates, deposited across the InP nanotemplate,
with narrow ($\sim$300~nm) gaps, allow in-plane electric fields to
be applied along the stripe (x-) direction and serve to isolate the
luminescence from individual QDs if the dot density is sufficiently
low.

\begin{figure}[]
  \centering
  \includegraphics[width=6.5cm]{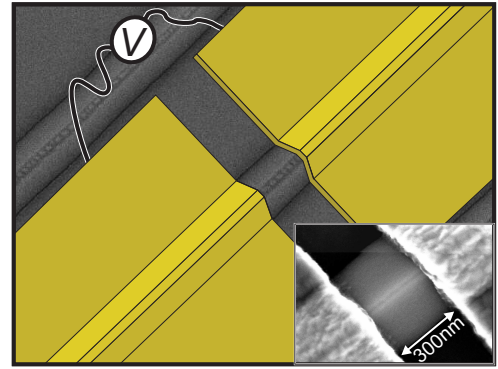}
  \caption{(Color Online) Schematic view of the device showing two metallic gates on top
of an SEM micrograph of an uncapped, "stripe geometry" InP ridge. A
linear array of InAs QDs is seen at the apex of the ridge,
in the space separating the gates. In real devices, the QDs are capped with InP prior to
gate deposition, as shown in the inset SEM micrograph, and the dot density
is chosen to be lower than that shown for the uncapped
sample, so that individual dots can be isolated.}\label{sample}
\end{figure}

Photoluminescence experiments were performed under applied electric
field, with an excitation intensity chosen to excite only the ground
state transition from the QD at zero applied electric field. Low
excitation intensity was required to ensure minimal screening of the
applied electric field.

To demonstrate the manipulation of the biexciton binding energy and
the engineering of Hidden Symmetry, we select an InAs/InP QD that
exhibits a negative \emph{XX} binding energy at zero applied bias.
Photoluminescence data from such a dot as a function of the applied
lateral electric field is shown in Fig. 2, with the appearance of
the \emph{XX} at lower energy than the \emph{X} shown in the inset,
at zero electric field and higher excitation power. At zero applied
bias, transitions from the two neutral exciton ground states, split
by e-h exchange, are observed at approximately 922.4~meV (1344~nm).
For biases from 0~V to approximately 1.9~V, the mean \emph{X} energy
shows a surprisingly small Stark shift, that will be discussed
subsequently, whilst the AES diminishes from 108~$\mu$eV to
56~$\mu$eV. Beyond 1.9~V, the AES is below the instrumental
resolution (50~$\mu$eV) and the oscillator strength of the \emph{X}
decays monotonically. Both the zero field value of the AES and its
reduction with the applied lateral field are consistent with the
behavior observed previously for InAs/GaAs QDs by Kowalik and
co-workers [15]. At approximately 1.9~V, the slowly red shifting
\emph{X} transition is crossed by a peak that blue shifts with
further increases in bias and a pair of peaks appear approximately 5
meV above the \emph{X}. This behavior is shown more clearly in Fig.
\ref{XXbinding}a, where the peak positions have been extracted from
Fig. 2. The general behavior discussed here is typical of single
InAs/InP QDs nucleated on "strip geometry" templates and has been
observed on eight separate QDs nucleated on templates with a variety
of widths. The observed behavior is generally insensitive to the
direction of the applied electric field, although the Schottky
contacts are sometimes "leaky" in one direction, preventing the
application of an electric field in this direction.

\begin{figure}[]
  \centering
  \includegraphics[width=8.3cm]{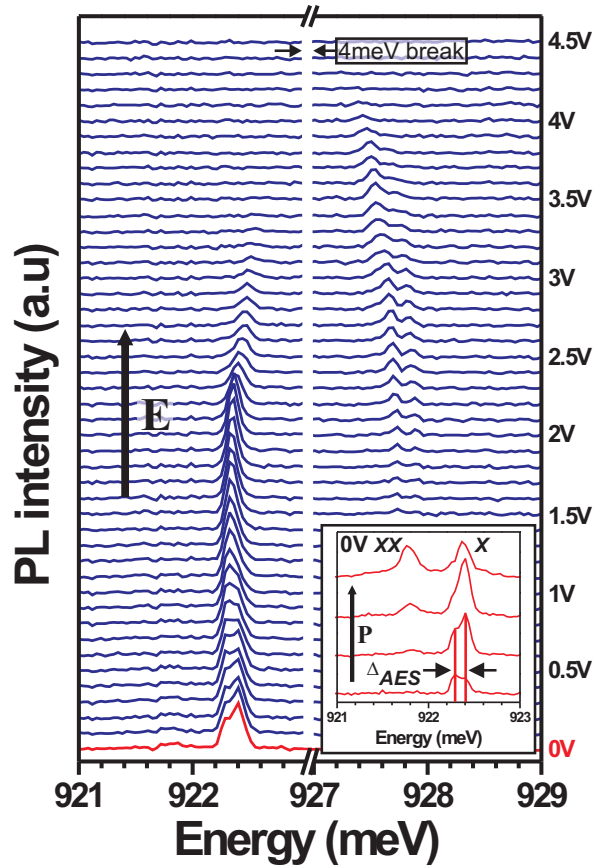}
  \caption{(Color Online) Electric field dependent PL spectra at T = 4.2 K.
There is a 4~meV break from 923~meV to 927~meV. (Inset) Power dependent PL
at 0~V. The biexciton (\emph{XX}), identified by its
quadratic power dependence, is observed at lower energy than the
exciton (\emph{X}) and has a binding energy of 510~$\mu$eV.} \label{AES}
\end{figure}

To understand the effects of the lateral electric field, we model
the QD using an isotropic, two-dimensional, parabolic confining
potential in the plane and calculate the expected emission energies
and intensities within the Full Configuration Interaction (CI)
approach. This involves constructing all possible configurations of
$N_{e}$ electrons and $N_{h}$ holes on the single-particle states in
the three lowest lying shells (\emph{s, p, d}), forming the
Hamiltonian matrix in the basis of these configurations, and
diagonalizing it numerically. Using the resulting excitonic
eigenenergies and eigenvalues, we calculate the emission spectrum
using Fermi's Golden Rule, $I(w)=\sum_{f}|\bra{f, N_{e}-1,
N_{h}-1}P^-\ket{i, N_{e}, N_{h}}|^2\delta(E_{i}-E_{f}-\hbar w)$,
where $P^{-}$ is the interband polarization operator removing one
electron-hole pair from the system,
$P^{-}=\sum_{ij}\alpha_{ij}c_{i}h_{j}$, and
$\alpha_{ij}={_{e}}\braket{i}{j}{_h}$  is the overlap integral,
controlling the optical selection rules for the emission. Here,
$c_{i}$, $h_{j}$ represent the the electron and hole annihilation
operators, respectively. The Hamiltonian utilized for the
interacting electrons and holes confined in the QD can be found in
ref \cite{Haw99}. In addition we add a term to the Hamiltonian that
describes the electron-hole exchange effects. In this work we use
the effective exchange Hamiltonian as described in Ref.
\cite{Bayer02}. The results of these calculations are shown in Fig.
\ref{XXbinding}b, for the ground and excited (forbidden) state of
the neutral exciton and for the ground state of the biexciton. At
zero applied electric field, the neutral exciton transitions
correspond to the recombination of a single electron and hole in the
s-shell of a parabolic QD. For an electric field $E$ applied along
the x axis, the confining potential in the x direction,
$V(x)=\frac{1}{2}m_{e}^*w_{e}^2x^2-eEx$, remains parabolic,
$V(x)=\frac{1}{2}m_{e}^*w_{e}^2(x-\Delta
x_e)^2-\Delta\varepsilon_{e}$, with a spatial shift $\Delta
x_{e}=+eE/m_{e}^*w_{e}^2$ for the electron and $\Delta
x_{h}=+eE/m_{h}^*w_{h}^2$ for the hole. Here $e$ and $m_{e}$ are the
electron charge and effective mass respectively and $\hbar w_{e}$ is
the characteristic energy of the confinement. The equivalent hole
parameters are labeled by the subscript $h$. The relative shift of
the origins of electron and hole confinements results in a change of
overlaps between electron and hole states, influencing the
oscillator strength of the transitions. In the calculations
presented here we assume the following system parameters:  $\hbar
w_{e}=12$~meV, $m_{e}^{*}= 0.055m_{0}$, $\hbar w_{h}=$ 6~meV, and
$m_{h}^{*}=0.11m_{0}$ with $m_{0}$ being the mass of a free
electron.  Under applied lateral field, the induced e-h charge
separation produces a reduction in the \emph{X} oscillator strength
and allows transitions between s-shell electrons and p-shell holes
($s-p_{x}$) that were symmetry forbidden at zero electric field. The
predicted behavior allows us to identify the two peaks appearing in
experiment under bias, approximately 5~meV above the \emph{X}, as
$s-p_{x}$ and $s-p_{y}$ transitions from a slightly asymmetric QD in
which the electric field does not coincide exactly with either of
the QD axes. For the \emph{X} transitions, the CI calculations
predict a monotonic decrease of the AES with applied electric field,
as observed experimentally, and a Stark shift that is finite, but
much smaller than that expected for a non-interacting electron-hole
pair. In addition to the red shifting \emph{X}, $s-p_{x}$ and
$s-p_{y}$ transitions, the CI calculations predict a blue shift in
the \emph{XX} transition with increasing electric field and an
eventual crossing of the \emph{X} and \emph{XX} transition energies.
At this crossing point the binding energy of the \emph{XX} vanishes
($B.E._{XX}=0$) and the QD displays Hidden Symmetry. In experiment,
the crossing of the \emph{X} and \emph{XX} is observed at
approximately 1.9~V and the \emph{XX} blue shift is observed above
this. The appearance of the \emph{XX} at the low excitation powers
used here to avoid screening effects, is somewhat surprising, but
might be expected due to the increased radiative lifetime of the
\emph{X} as charge separates under the applied field
\cite{Stavarache06}.

\begin{figure}[]
  \centering
  \includegraphics[width=7cm]{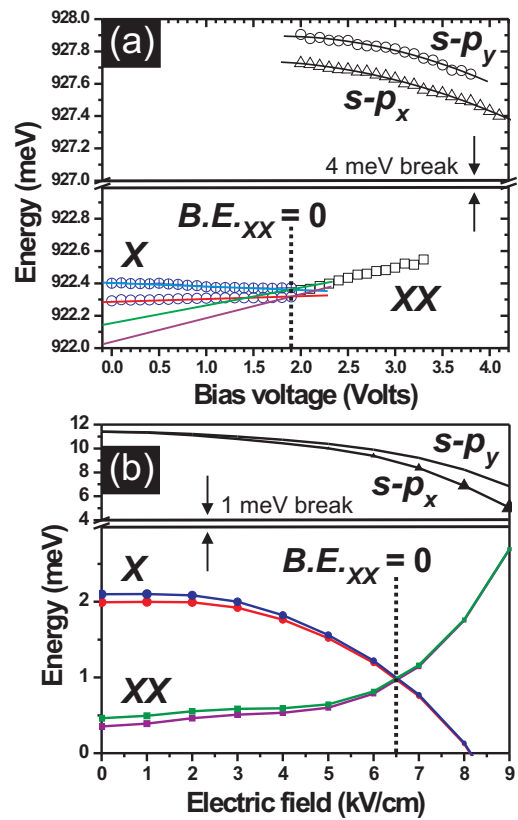}
  \caption{(Color Online) Tuning of the biexciton binding energy to zero ($B.E._{XX}=0$). (a)
Peak energies extracted from Fig. 2 as a function of the electric
field (bias voltage). The AES splitting of the exciton, \emph{X}, (open blue
circles) diminishes with increasing bias until the biexciton, \emph{XX},
(open black squares) appears at approximately 1.9~V. Forbidden transitions, $s-p_{x}$ (open black
triangles) and $s-p_{y}$ (open black circles) are shown with a quadratic
fit (solid black line). Transitions from Fig. 4 are shown
schematically with the experimental data. (b)
Calculated PL transitions as a function of lateral electric field
obtained from Fermi's golden rule. Symbols size (x10 for $s-p_{x}$, $s-p_{y}$) is
used to indicate the size of the e-h overlap.  Exciton, \emph{X}, and
biexciton, \emph{XX}, transitions ($B_{1}$, $A_{1}$, $A_{2}$, $B_{2}$) are colour coded in the same manner used
for Fig. 4.}\label{XXbinding}
\end{figure}

For a source of entangled photon pairs, the benefit of voltage
induced Hidden Symmetry is shown in Fig 4. Within existing
proposals, the two channels for biexciton decay ($A$ and $B$) are
made indistinguishable by setting  $\Delta_{AES}=0$, so that
transitions $A_{1}$ and $B_{1}$ in Fig. 4a become degenerate, as
well as the transitions $A_{2}$ and $B_{2}$. Removal of the AES for
arrays of individual QDs is technologically difficult however since
the application of large magnetic fields \cite{Stevenson06} or
compositional modifications \cite{Gre06} cannot be done locally on a
dot to dot basis. However, considering Fig. 4b, it is clear that in
the presence of Hidden symmetry, where
$\Delta{E_{XX}}=\Delta{E_{X}}$, we may identify two pairs of
degenerate transitions even in the presence of a non-zero AES:
$A_{1}$ and $B_{2}$, as well as $A_{2}$ and $B_{1}$. This behavior
restores the indistinguishability of the biexciton decay channels
and suggests that arrays of individually addressable, voltage
controlled QDs would be able to provide multiple sources of
entangled photons, in which the two-photon wave function is
entangled in the horizontal ($H$) and vertical ($V$) polarisation
degrees of freedom, $\psi=\frac{1}{\sqrt{2}}(\ket{HH}+\ket{VV})$.
Within the proposed scheme it would be possible, in principle, to
distinguish the decay channel if temporal information were
available. However, this objection is unfounded since a
combination of linear optical elements following the QD, such as
polarizing beam splitters, energy filters, and optical delay lines
can be used to completely remove any timing information and restore
temporal overlap. This removal of the timing information is facilitated by the separation into horizontal and vertical polarizations for both photons on each arm of the $XX$ decay.

\begin{figure}[]
  \centering
  \includegraphics[width=7cm]{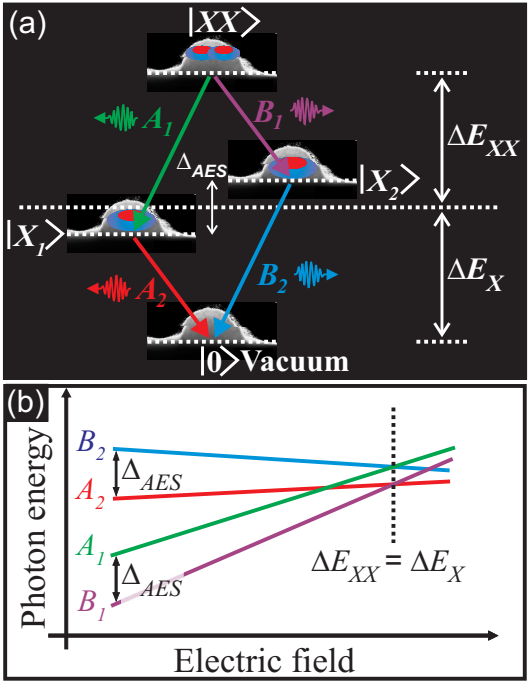}
  \caption{(Color Online) Polarization entangled photon source exhibiting exchange
splitting. (a) Schematic diagram of the biexciton-radiative cascade
for a polarization entangled photon source exhibiting an anisotropic
exchange splitting (AES) when the binding energy of the biexciton is
tuned to zero ($\Delta E_{xx}=\Delta E_{x}$). Two possible decay paths
exist from the biexciton $\ket{XX}$ to vacuum $\ket{0}$ through the intermediate,
non-degenerate excitonic states $\ket{X_{1}}$ and $\ket{X_{2}}$. The solid lines with
arrows indicate transitions with energy $A_{1}$ (green), $A_{2}$
(red), $B_{1}$ (purple) and $B_{2}$ (blue). The two possible decay
paths of the $XX-X$ cascade are represented by $A$ and $B$, whilst
the subscript 1 (2) denotes the first (second) photon emitted in the
cascade. (b) Schematic view of transitions in (a) as a function of
lateral electric field with the same colour scheme for transitions.
For the case when $\Delta E_{xx}=\Delta E_{x}$, the biexciton binding
energy vanishes and a polarization entangled photon pair is
possible, even in the presence of exchange splitting.}\label{EPP}
\end{figure}

For such a device to operate, the engineered crossing of the $X$ and
$XX$ lines is critical and can be easily understood. In the absence
of e-h exchange, the two optically active exciton states have the
same energy,
$E_{X}=\varepsilon_{00}^{e}+\varepsilon_{00}^{h}-V_{eh}^{0}-
\Delta{E_{X}^{CORR}}$, corresponding to the sum of two
single-particle energies,
$\varepsilon_{00}^{e}+\varepsilon_{00}^{h}$, an e-h attraction,
$-V_{eh}^{0}$, and the correlation energy, $\Delta{E_{X}^{CORR}}$.
The e-h exchange mixes the two exciton configurations and splits
their energies; producing two excitonic states with energies
$E_{X1,X2}=\varepsilon_{00}^{e}+\varepsilon_{00}^{h}-V_{eh}^{0}-
\Delta{E_{X}^{CORR}}\mp\frac{\Delta_{AES}}{2}$. These are the
intermediate states in the biexciton cascade, shown schematically in
Fig. 4a. Since the final state in the excitonic recombination is the
vacuum, the above energies also define the PL peak positions
corresponding to the transitions $A_{2}$ and $B_{2}$. As the
electrostatic field $E$ is increased, both single-particle terms
decrease due to the Stark effect. However, detailed calculations
\cite{Kor07} show that $V_{eh}^0$ and the correlation energy
$\Delta{E_{X}^{CORR}}$ also decrease with increasing electric field and since they appear with a
negative sign, changes in the single-particle energies are
compensated. This behavior is the explanation for the small Stark shift
observed in Fig. 2. In a similar manner, the energy of the biexciton
can be approximated by,
$E_{XX}=2\varepsilon_{00}^{e}+2\varepsilon_{00}^{h}+V_{ee}^{0}
+V_{hh}^0-4V_{eh}^{0}-\Delta{E_{X}^{CORR}}$. The energies of photons
emitted in transitions $A_{1}$ and $B_{1}$ correspond to the
differences between this energy and those of the two single
excitons:
$E_{A1,B1}=(\varepsilon_{00}^{e}+\varepsilon_{00}^{h}-V_{eh}^{0}-
\Delta{E_{X}^{CORR}}\pm\frac{\Delta_{AES}}{2})
+(V_{ee}^0+V_{hh}^0-2V_{eh}^0)
-(\Delta{E_{X}^{CORR}}-2\Delta{E_{XX}^{CORR}})$. In this expression,
the first term is identical to the PL peak positions of the single
excitons, only with opposite contribution from the AES,  and the second
term accounts for the Coulomb interactions between the two
electron-hole pairs building the biexciton. At zero electric field
this second term is approximately zero, since the repulsion between
carriers of the same type is balanced by electron-hole attraction, and
it is the third term, comparing the correlation corrections of
the biexciton to those of the exciton, which defines the splitting
between the PL lines of the two complexes. This third term is
usually negative, resulting in biexcitonic PL peaks that appear at
lower energies than those of single excitons. With increasing
electric field, the first term is almost unchanged because of the
competing single particle and electron-hole Coulomb terms.  However,
as the electrons and holes are separated, the electron-hole
attraction decreases, driving the biexciton PL peak positions
towards higher energies. For a sufficiently large electric field the
interaction terms exactly cancel the correlation corrections,
resulting in a degeneracy of the biexciton and exciton emission and
allowing indistinguishable biexciton decay channels to be
identified.

The authors would like to acknowledge the financial support of the
Canadian Institute for Advanced Research, the Canadian Institute for
Photonic Innovations, QuantumWorks and the Natural Sciences and Engineering
Research Council.

\end{document}